\documentclass{aa}  

\usepackage[T1]{fontenc}

\usepackage{graphicx}
\usepackage{multirow}
\usepackage{txfonts}
\usepackage{enumitem}
\usepackage{subcaption}
\usepackage{hyperref}

\newcommand{\orcid}[1]{\href{https://orcid.org/#1}{\includegraphics[width=8pt]{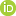}}}

\begin{document}

   \title{Tuning into spatial frequency space}

   \subtitle{Satellite and space debris detection in the ZTF alert stream.}

   \author{
        J. P. Carvajal  \inst{1,2}\fnmsep\thanks{Corresponding author: Jcarvajal000@gmail.com}\orcid{0000-0001-6584-7104}
            \and
        F. E. Bauer\inst{3}\orcid{0000-0002-8686-8737}
            \and 
        I. Reyes-Jainaga\inst{4}\orcid{0000-0003-3627-0216} 
            \and 
        F. Förster \inst{2,5,6}\orcid{0000-0003-3459-2270}
            \and 
        A. M. Mu\~noz Arancibia\inst{2,5}\orcid{0000-0002-8722-516X}
            \and
        M. Catelan \inst{1,2,7}\orcid{0000-0001-6003-8877}
            \and
        P. Sánchez-Sáez \inst{8,2}\orcid{https://orcid.org/0000-0003-0820-4692}
            \and
        C. Ricci \inst{9,10}\orcid{0000-0001-5231-2645}
            \and
        A. Bayo \inst{8}\orcid{0000-0001-7868-7031}
          }

    \institute{
    Instituto de Astrofísica, Pontificia Universidad Católica de Chile, Av. Vicuña Mackenna 4860, 7820436 Macul, Santiago, Chile  
         \and
    Millennium Institute of Astrophysics (MAS), Nuncio Monsenor Sótero Sanz 100, Providencia, Santiago, Chile
        \and
    Instituto de Alta Investigaci{\'{o}}n, Universidad de Tarapac{\'{a}}, Casilla 7D, Arica, Chile
        \and
    Data Observatory, Av. Eliodoro Yáñez 2990, oficina A5, Providencia, Chile
        \and
    Center for Mathematical Modeling (CMM), University of Chile, AFB170001, Santiago, Chile
        \and
    Data \& Artificial Intelligence Initiative (ID\&IA), University of Chile, Santiago, Chile
        \and
    Centro de Astro-Ingeniería, Pontificia Universidad Católica de Chile, Av. Vicuña Mackenna 4860, 7820436 Macul, Santiago, Chile
        \and
    European Southern Observatory, Karl-Schwarzschild-Strasse 2, 85748 Garching bei München, Germany
        \and
    Instituto de Estudios Astrofísicos, Facultad de Ingeniería y Ciencias, Universidad Diego Portales, Av. Ejército Libertador 441, Santiago, Chile
        \and
    Kavli Institute for Astronomy and Astrophysics, Peking University, Beijing 100871, China
        }

   \date{Received September 15, 1996; accepted March 16, 1997}
 
  \abstract
   {A significant challenge in the study of transient astrophysical phenomena is the identification of bogus events, among which human-made Earth-orbiting satellites and debris remain major contaminants. Existing pipelines effectively identify satellite trails but can miss more complex signatures, such as collections of satellite glints. In the Rubin Observatory era, the scale of operations will increase tenfold with respect to its precursor, the Zwicky Transient Facility (ZTF), requiring crucial improvements in classification purity, data compression for informative alerts, and pipeline speed.}
   {We explore the use of 2D Fast Fourier Transform (FFT) on difference images as a tool to improve satellite-detection machine learning algorithms.
   }
   {Using the Automatic Learning for the Rapid Classification of Events (ALeRCE) single-stamp classifier as a baseline, we adapt its architecture to receive a cutout of the FFT of the difference image, in addition to the three (science, reference, difference) ZTF image cutouts (hereafter stamps).
   {We explore various stamp sizes and resolutions, assessing the benefits of incorporating FFT images, particularly when data compression is critical due to alert size limitations and pipeline speed constraints (e.g., in large-scale surveys such as the Legacy Survey of Space and Time).}
   }
   {{The inclusion of the FFT can significantly improve satellite detection performance. The most notable improvement occurred in the smallest field-of-view model (16$\arcsec$), whose satellite classification accuracy increased from $(72.0\pm2.9)\%$ to $(87.8\pm1.3)\%$ after including the FFT, computed from the full 63$\arcsec$ difference images. This demonstrates the effectiveness of FFT in compressing and extracting relevant large-scale satellite features. However, the FFT alone did not fully match the accuracy achieved by the full 63$\arcsec$, $(95.9\pm1.3)\%$ and multiscale $(90.6\pm0.8)\%$ models, highlighting the complementary importance of contextual spatial information.}}
   {We show how FFTs can be leveraged to cull satellite and space debris signatures from alert streams.}

   \keywords{Techniques: image processing -- Techniques: photometric -- Methods: data analysis -- Surveys.}
   \maketitle

\section{Introduction}
\label{sec:intro}

\begin{figure*}[hptb]
    \centering
    \includegraphics[width=0.9\textwidth]{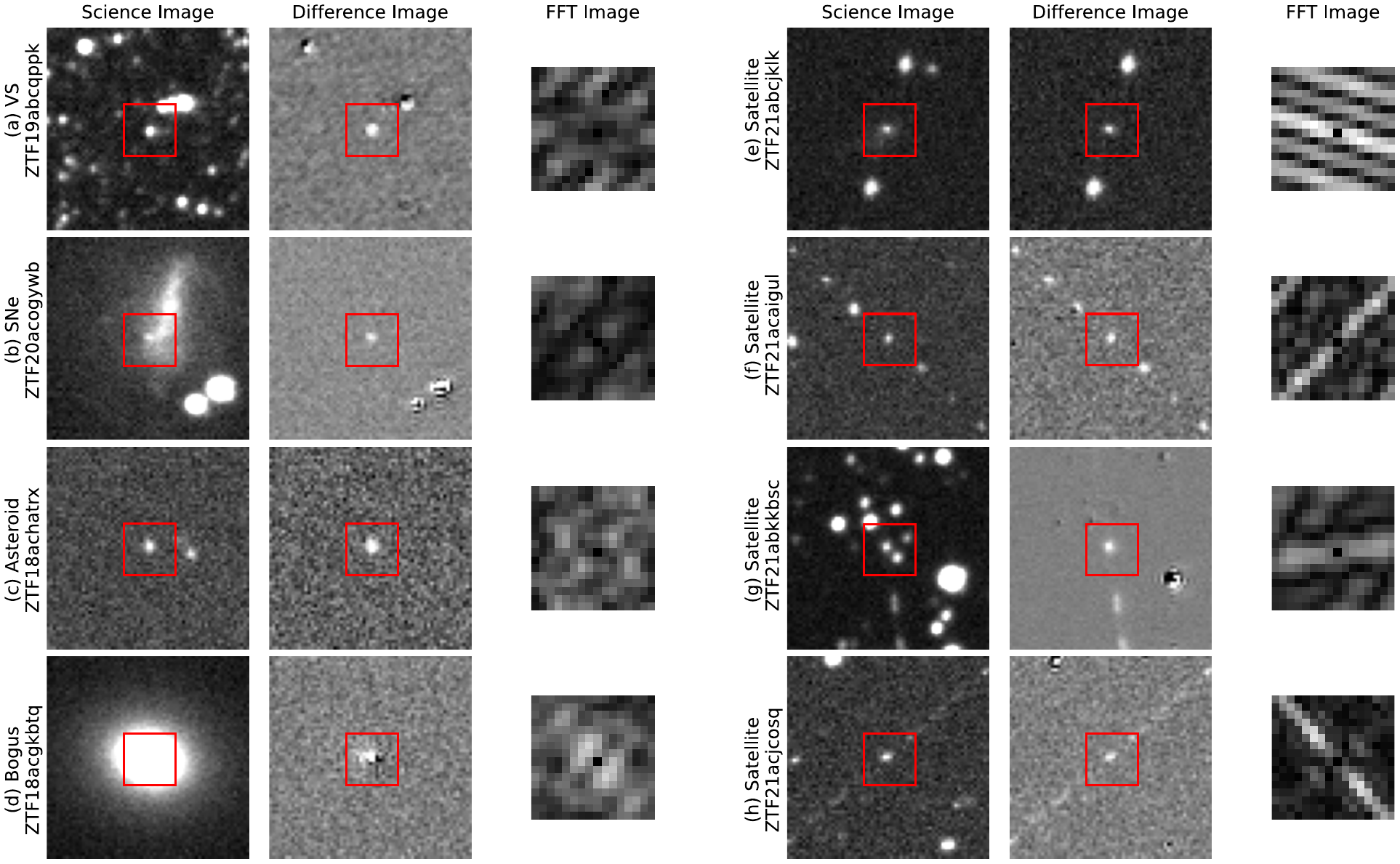}
    \caption{
    Comparison of science image, difference image, and its FFT for various transient sources and contaminants in the ZTF alert stream. The full 63$\arcsec$~$\times$~63$\arcsec$ science and difference stamps are shown, with the central 16$\arcsec$~$\times$~16$\arcsec$ denoted in red. The FFT image stamp is cropped into the central 16~$\times$~16 pixels. Panels (a)-(c) display astrophysical events: (a) variable star ZTF19abcqppk; (b) supernova ZTF20acogywb; and (c) asteroid ZTF18achatrx. Panel (d) shows a typical bogus event (ZTF18acgkbtq), characteristic of a bad subtraction. Panels (e)-(h) highlight different types of satellite signatures that reached the alert stream: (e) and (f), from alerts ZTF21abcjklk and ZTF21acaigul, show the typical regular glint signature with different separation among glints; (g) shows a satellite with irregular glints and an overall asymmetric signature; (h) shows a satellite with both a continuous signature and glints. This small portion of Fourier space effectively captures the distinct extended signals of satellites.}
    \label{fig:Sat_signature}
\end{figure*}

Large étendue telescopes engaging in time-domain surveys, such as the Zwicky Transient Facility \citep[ZTF,][]{Bellm_2019_ZTF, Graham_2019_ZTF}, have quickly become a cornerstone of observational astrophysics. ZTF has revolutionized our ability to detect and characterize a wide variety of transients, active galaxies, and variable star phenomena \citep[e.g.,][]{Chen_2020_Vstars_ZTF, Perley_2020_ZTF_SNe, Carrasco-Davis_2021_ALeRCE, SanchezSaez_2021_Anomaly, SanchezSaez_2023_ZTF_AGN}, through both its 5-$\sigma$ difference-image alert stream and frequent data releases \citep{Masci_2019_ZTF_pipeline}. 
Overall, ZTF distributes up to $10^{6}$ alerts each night (with a median closer to $\sim$200\,000) which equates to tens of GB of data per night \citep{Patterson_2019_ZTF_ADS}. The expected rates for the Legacy Survey of Space and Time \citep[LSST,][]{LSST_ScienceBook2009, Ivezic_2019_LSST} from the Vera C. Rubin Observatory will be an order of magnitude larger, with a single-visit depth almost 4 magnitudes deeper in the $r$ band compared to ZTF. A handful of so-called community alert brokers have dedicated themselves to the distribution, filtering, and annotation of these massive alert volumes. These brokers ingest the alert streams and provide processed science products and services that enable the community to pursue their science goals without overloading the observatories' infrastructure. Current full-stream community brokers for ZTF and future LSST include 
ALeRCE \citep{Forster_2021_alerce}, 
AMPEL \citep{Nordin_2019_AMPEL}, 
ANTARES \citep{Matheson_2021_ANTARES}, 
Babamul, 
Fink \citep{Moller_2020_FINK}, 
Lasair \citep{Smith_2019_Lasair}, and 
Pitt-Google\footnote{\href{https://pitt-broker.readthedocs.io/en/latest/}{https://pitt-broker.readthedocs.io/en/latest/}}.

One of the persistent challenges is the identification and removal of contaminants (often referred to as ``bogus'' detections) from the observations and the transient event alert stream. Mitigation of bogus events is particularly important for robust early transient discovery and classification, such as identifying flash-ionization episodes in young supernovae (SNe) or short-lived kilonovae; this was a prime motivation for the development of the real-time first-stamp classifier by ALeRCE \citep[][hereafter, the ``Stamp Classifier'']{Carrasco-Davis_2021_ALeRCE}. Notably, human-made satellites and space debris orbiting Earth can be particularly complicated contaminants to identify in real-time for existing algorithms and pipelines. \citet{Karpov_2022_satellite_glints, Karpov_2023_satellite_glints} quantified the effect of satellite glints (see Fig.~\ref{fig:Sat_signature}) that bypassed both the streak masking algorithm \citep{Laher_2014_streak_mask} and real-bogus classifier \citep{Duev_2019_ZTF_bogus} components of ZTF's pipeline \citep{Masci_2019_ZTF_pipeline}, finding at least $\approx$3000 glint-related alerts per month. For reference, this is $\approx$4.4 times the monthly rate of new transients detected by ZTF and reported to the Transient Name Server (TNS)\footnote{\href{https://www.wis-tns.org/}{https://www.wis-tns.org/}}.

This challenge is already a significant in the ZTF real-time alert stream and stands to become even more problematic in the LSST era due to the shift in stamp field of view (FoV), which is slated to go from 63$\arcsec$~$\times$~63$\arcsec$ at a 1$\arcsec/\text{pixel}$ resolution for ZTF (e.g., Fig.~\ref{fig:Sat_signature}) to 6$\arcsec$~$\times$~6$\arcsec$ at a 0.2$\arcsec/\text{pixel}$ resolution for LSST, due to bandwidth limitations associated with increased image resolution. 
An upside of the increased resolution is highlighted in recent work by \citet{Tyson_2024}, which shows how glints from tumbling low-Earth orbit debris may become identifiable due to distinct de-focus and shape characteristics. This, however, does not completely alleviate the concerns, as glint-producing satellites occupy a wide range of orbits \citep{Karpov_2023_satellite_glints}. 
As discussed in \citet[][hereafter RJ23]{Reyes-Jainaga_2023_alerce_multistamp}, a promising approach to mitigate the FoV reduction in LSST is to employ multiscale image stamps. These stamps capture a larger FoV with decreasing spatial resolution as the distance from the source center increases, effectively compressing data by preserving high-detail information in the central region while also retaining relevant contextual information, such as flux from satellite trails, around the alert.

Motivated by the {speed and} potential for image compression{,} and by the unique glint signatures of satellites and space debris (hereafter, simply "satellites"), we explore the spatial frequency domain of the difference images via the Fast Fourier Transform (FFT) algorithm \citep{cooley1965FFT}. 
Satellite glints, presumably caused by rotating objects that reflect sunlight, exhibit a range of patterns from simple (e.g., Fig.~\ref{fig:Sat_signature}, panel f) to complex (e.g., Fig.~\ref{fig:Sat_signature}, panel g).
Additionally, multiple glints from a single satellite may not be captured within the FoV of a single stamp (e.g., see Sect.~\ref{sec:discussionBeyond}).
These glints generally consist of a periodic signal, dominated by the satellite’s rotation, aligned along an approximately straight line within the locality of the image. The signal is convolved with the Point Spread Function (PSF), and under this locality approximation, we can assume a relatively constant illumination geometry. Therefore, inspecting the spatial frequencies of these satellite glint variations is a natural path to explore.
Central, point-like emission maps to very extended scales in Fourier space, while extended emission maps to more compact scales. 
Thus, in contrast with the multiscale approach of RJ23, the central portions of the Fourier map naturally tap into the extended emission, while imaging cropping effectively limits the minimum scales probed, rather than the maximum ones.

Building upon the approach in RJ23, we analyze the effect of including FFT information within the adapted architecture of the ALeRCE Stamp Classifier. The incorporation of the FFT slightly increases the model's input size\footnote{For example, for a central 16~$\times$~16~px portion, this results in 128 floats (16~$\times$~16~$\div$~2, due to the complex conjugate symmetry of the FFT).}, but {importantly, the efficiency of existing FFT implementations ensures that this addition incurs only a small computational cost, even in large-scale applications (see Appendix~\ref{sec:benchmarking}).} We aim to assess its potential for capturing extended spatial features in a compact form.
We present the methods and results in Sects.~\ref{sec:methods} and \ref{sec:results}, respectively. In Sect.~\ref{sec:discussion} we analyze the results and discuss the future prospects for the FFT, and how it can efficiently expand opportunities for improving streak detection algorithms \citep[e.g.,][]{Duev_2019_streaks} to perform better on satellite glints while keeping inputs small, which will be fundamental in the LSST era both for internal pipelines and downstream filtering.

\section{Enhancing the ALeRCE Stamp Classifier with FFT}
\label{sec:methods}

The ALeRCE Stamp Classifier was first introduced in \citet{Carrasco-Davis_2021_ALeRCE}, adopting a Convolutional Neural Network (CNN) model that classifies individual ZTF alerts based on three-channel image cutouts (science, reference, and difference) and additional metadata (e.g., coordinates, flux and magnitude measurements, and seeing) into four distinct astrophysical sources and bogus. The effectiveness of this machine learning (ML) model is demonstrated by ALeRCE's position as one of the top three reporting groups of transients to TNS, with $\lesssim$1\% contamination among spectroscopically classified transients first reported by \citet{Forster_2021_alerce}. 

In this work, we explore the effect of incorporating FFTs in the Stamp Classifier, building upon previous works. Our modifications start from the model architecture introduced in RJ23, where the Satellite class was first introduced. The classifier categorizes alerts into six distinct classes: supernovae (SNe), variable stars, asteroids, active galactic nuclei (AGN), satellites, and bogus detections.

\subsection{Preprocessing}

After the ingestion of each alert, a preprocessing step takes place before the model run.
A schematic is shown in Fig.~\ref{fig:preproc_diag}. The detailed steps to go from the base ZTF stamps to the cropped stamp cube (blue box in the schematic), that serves as input for the model, are the following:

\begin{figure}[htpb]
    \centering
    \includegraphics[width=0.9\linewidth]{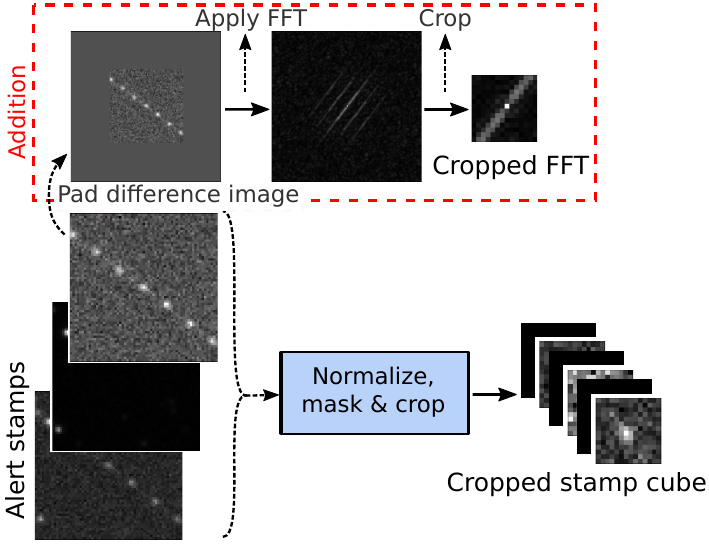}
    \caption{Schematic of the preprocessing for our FFT-enhanced Stamp Classifier. The red dashed box encloses the new steps for the inclusion of the FFT with respect to previous versions of the classifier (e.g., RJ23).}
    \label{fig:preproc_diag}
\end{figure}

\begin{enumerate}
    \item In the rare cases where the stamps do not have the original dimensions of 63~$\times$~63 pixels (e.g., the object is by the edge of the CCD), the alert is dropped. The current necessity for the source to lie in the center of the stamp complicates such modifications. An attempt to ``repair'' these cases has yet to be implemented. 
    \item In each of the stamps, a boolean integrity mask is constructed, where infinite and NaN (Not a Number) values in the stamp are associated with a True (1), and finite values with a False (0).
    \item The non-finite values in the masked stamp are then set to zero (0).
    \item For computational {efficiency}, the normalization is performed {separately for each stamp type (science, reference, difference) in batches}. The maximum for the normalization is the 99th percentile of the absolute values. Then, the values are clipped to 2.0.
    \item The normalized stamps are then concatenated with the integrity mask, ending up with a 6-channel image. When employing the Multiscale model, this process is repeated across four different scales, resulting in a 24-channel image.
    \item Ultimately, the stamps are cropped or resized, depending on the model they will go into.
\end{enumerate}

Beyond the preprocessing steps used in RJ23, we introduce new steps (highlighted in the red dashed box in Fig.~\ref{fig:preproc_diag}) to incorporate the FFT calculation. Specifically, these steps are:

\begin{enumerate}[resume]
    \item Starting from the complete (63~$\times$~63 pixels) normalized difference image from step 5, we pad it to 128 pixels, filling with the median. {The selection of the size was based on the efficiency of the FFT algorithm for sizes that are powers of two, and additional considerations explained below}.
    \item We then apply the two-dimensional FFT and take the norm to obtain the amplitude.
    \item Finally, we crop the central portion of the FFT. In our experiments, we kept the central 16~$\times$~16~px region in all cases.
\end{enumerate}

There are two reasons to apply the padding, and both are associated with the nature of the FFT algorithm. First, since the transform maintains the sampling, padding allows increasing the spatial frequency resolution. Second, the FFT algorithm is constructed under the assumption that the sequence to be transformed is infinitely periodic. Since this is not the case for these images, artifacts that might appear due to this inconsistency are mitigated by the padding. {For these two reasons ---achieving higher spatial frequency resolution and mitigating periodicity artifacts--- the stamps are padded to 128 pixels instead of just 64. Increasing the padding beyond 128 pixels did not yield improvements in our early tests, and would require larger cutouts to achieve the same coverage of spatial frequencies.} Similar measures are taken in other fields that use the FFT, such as computational Fourier optics, to deal with such issues \citep[e.g.,][]{Voelz2011-CompFourierOptics}.
Although the FFT is a complex field, we focus on the amplitude due to the potential for data compression. Alternatives to this are discussed in Sect.~\ref{sec:discussionBeyond}.

\subsection{Models}
\label{sec:models}

\begin{figure*}[hptb]
    \centering
    \includegraphics[width=0.9\textwidth]{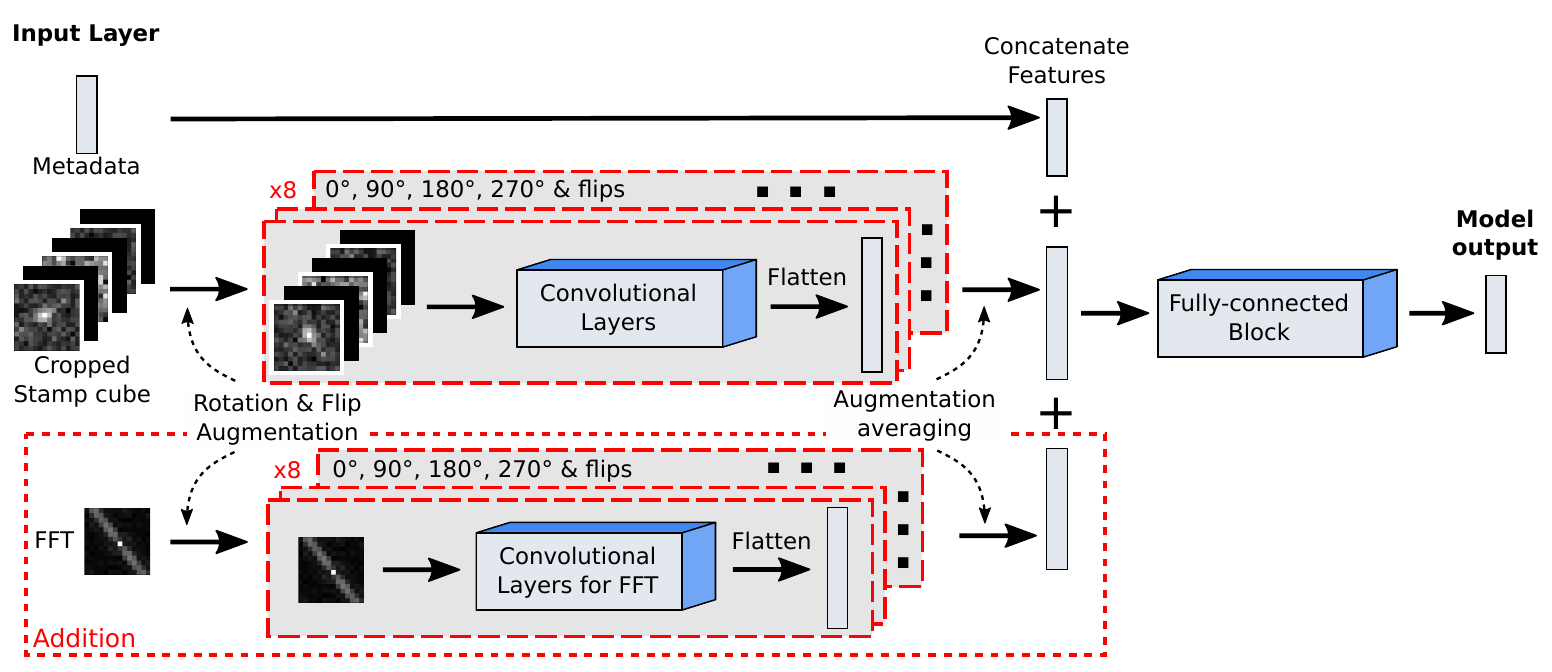}
    \caption{A schematic view of the architecture of the ML models used in this work. The input includes the cropped stamp cube, the FFT cutout, and additional metadata. Both the stamp cube and FFT cutout undergo rotation and flip augmentation before being processed through separate convolutional blocks (red long-dashed boxes). The outputs are then concatenated with the metadata and passed through a fully-connected block for prediction. The red short-dashed box highlights the addition of the FFT block to the architecture, which was included in models D, E, and F (see Sect.~\ref{sec:models}).}
    \label{fig:model}
\end{figure*}

Fig.~\ref{fig:model} illustrates the architecture of the models evaluated in this study. The key components include the cropped stamp cube, the FFT cutout from the preprocessing step, and additional metadata (primarily the coordinates of the alert). These inputs are processed through distinct neural network blocks before being combined in a final fully-connected layer that produces the model predictions.

In the architecture, the cropped stamp cube is passed through a convolutional block (referred to as the Stamp convolutional block) that applies rotational and flip augmentations. After the convolutional layers, the images are flattened, averaged over the augmentation steps, and passed through a dropout layer before being concatenated with the other features.
While this block mirrors the structural design from RJ23, for the models that feature the FFT as an input (see the red short-dashed box in Fig.~\ref{fig:model}), the hyperparameters were re-optimized separately. 

Parallel to this, the FFT convolutional block processes the FFT cutout. Similar to the Stamp convolutional block, it includes rotation and flip augmentation followed by convolutional and pooling layers, and finally, the output is flattened and concatenated with the other features.

To assess the value of incorporating the FFT block, we compared models that included it with those that did not:

\begin{enumerate}[label=\Alph*., leftmargin=2em, align=left]
    \item \textbf{Full}: The full 63$\arcsec$~$\times$~63$\arcsec$ (63~$\times$~63~px) FoV alert at 1$\arcsec$ resolution is used. The input consists of three images (science, reference, and difference) and their corresponding integrity masks, creating a six-channel image.
    \item \textbf{Multiscale}: The stamp cube includes 8$\arcsec$, 16$\arcsec$, 32$\arcsec$, and 63$\arcsec$ FoV stamps at corresponding resolutions of 1$\arcsec$, 2$\arcsec$, 4$\arcsec$, and 8$\arcsec$. This produces a 24-channel image with four channels per image type.
    \item \textbf{Cropped-16}: The alert is cropped to a 16$\arcsec$ FoV with the full 1$\arcsec$ resolution.
    \item \textbf{Full + FFT}: Combines the full 63$\arcsec$ FoV with the FFT block.
    \item \textbf{Multiscale + FFT}: Combines the multiscale stamp cube with the FFT block.
    \item \textbf{Cropped-16 + FFT}: Combines the cropped 16$\arcsec$ FoV with the FFT block.
\end{enumerate}

Model F (Cropped-16 + FFT) is particularly interesting due to its scalability. The amount of information in an image scales as the square of the number of pixels ($N^2$). However, the FFT algorithm's computational cost scales as $N*\log(N)$, making it significantly efficient, especially considering the subsequent cropping of the FFT to retain only the {most relevant information}. This efficiency allows model F to capture large-scale features of transient phenomena, such as satellite glints, while minimizing the inclusion of non-informative data (e.g., sky pixels in a science image) that would slow the pipelines. 

For completeness, models D and E were also evaluated, although they did not show statistically significant improvements compared to models A and B. This is because the FFT block provided redundant information that was already captured by the larger FoVs in these models. The inclusion of the FFT block in these cases added complexity to the models, making them slower {and complicating their optimization}.

\subsection{Data and training}
\label{sec:dataAndTraining}

{We started from the same dataset used by RJ23, and closely followed the training procedure outlined therein. A detailed description of the dataset and its construction (by cross-matching with other databases) for all classes except Satellites can be found in }\citet{Carrasco-Davis_2021_ALeRCE} and \citet{SanchezSaez_2021_LC_Classifier}. {Since RJ23, in the case of Satellites, most of the alerts were flagged during the human inspection of ZTF alerts while searching for SNe candidates. While this approach limits the number of objects available, it ensures the purity of our sample, as multiple glints or a trail-like signal within the 63$\arcsec\times$63$\arcsec$ FoV of the stamps are necessary to confidently classify the object as a satellite with only the alert information. One fundamental revision was made in this work. Inspection of  alerts confused among previous models prompted a manual review of all alerts labeled as bogus in the dataset. We found that a significant portion ($\sim3$\%) of the bogus-labeled alerts showed clear satellite trails and glints. Thus, we re-labeled them as satellites. We also added more satellites flagged during 2023-2024, increasing our sample of satellites to 1204. This manual revision significantly reduced contamination and improved the representativeness of our satellite class, thereby enhancing model reliability. Table~\ref{tab:TrainingSet} summarizes the number of alerts per class. To deal with the remaining class imbalance, we employed balanced training batches and class-balanced cross-entropy loss for validation (see RJ23 for details). A limitation of this approach is discussed in Sect.~\ref{sec:discussionFFT}.}

\begin{table}
\caption{Number of objects per class.}
\label{tab:TrainingSet}
\centering
\begin{tabular}{c c c c c c}
\hline
AGN & Asteroid & Bogus & Satellite & SNe & VS \\
9,774 & 9,180 & 15,326 & 1204 & 3,615 & 10,211 \\
\hline
\end{tabular}
\end{table}

The training process involved a hyperparameter (HP) search using \texttt{Ray Tune} \citep{Liaw_2018_raytune}, optimizing parameters such as the number of filters in the convolutional layers, learning rate, dropout rate, the size of the first convolutional kernel, and the size of the last dense layer.

{Repeating the HP optimization for models A, B, and C after addressing bogus contamination resulted in moderate performance improvements, highlighting the value of this revision.} For models D, E, and F (with the FFT block), {we also ran a HP search}, focusing on the FFT-specific convolutional layers. The search space for inherited parameters was narrowed to one-third of the original range, centered around the optimal values from models A, B, and C, while the FFT-specific parameters were searched within a broad range of values.

This approach ensured efficient HP optimization for the FFT models, with search times comparable to the non-FFT models. After identifying the optimal hyperparameters, each model was trained five times using the entire training set to compute the F1 score, precision, and recall \citep[see Section 5.7.2. in][for definitions]{murphy2012_ML}, as well as to estimate their respective standard deviations and assess the significance of changes. The results are summarized in Table~\ref{tab:Metrics} for the test dataset.

\section{Results}
\label{sec:results}

\begin{table}
\caption{Performance metrics for each model}
\label{tab:Metrics}
\centering
\begin{tabular}{lcccc}
\hline\hline
Label & Model & F1 score & Precision & Recall \\
 & & (\%) & (\%) & (\%) \\
\hline
\multirow{2}{*}{A} & \multirow{2}{*}{Full} & 91.0 & 90.1 & 92.1 \\
 & & (0.7) & (1.2) & (0.3) \\
\hline
\multirow{2}{*}{B} & \multirow{2}{*}{Multiscale} & 91.1 & 90.5 & 91.7 \\
 & & (0.3) & (0.3) & (0.3) \\
\hline
\multirow{2}{*}{C} & \multirow{2}{*}{Cropped-16} & 87.9 & 88.0 & 87.9 \\
 & & (0.4) & (0.7) & (0.5) \\
\hline
\multirow{2}{*}{D} & \multirow{2}{*}{Full + FFT} & 91.0 & 90.7 & 91.4 \\
 & & (0.4) & (1.1) & (0.4) \\
\hline
\multirow{2}{*}{E} & \multirow{2}{*}{Multiscale + FFT} & 90.5 & 90.6 & 90.4 \\
 & & (0.3) & (0.4) & (0.4) \\
\hline
\multirow{2}{*}{F} & \multirow{2}{*}{Cropped-16 + FFT} & 90.9 & 91.1 & 90.8 \\
 & & (0.3) & (0.7) & (0.2) \\
\hline
\end{tabular}
\tablefoot{Mean values and standard deviations (in parentheses) are based on five training runs for each model.}
\end{table}

{In this section we briefly introduce the model performance metrics, while interpretation follows in Sect.~\ref{sec:discussion}. }Table~\ref{tab:Metrics} presents the performance metrics for each model. As in RJ23, the Multiscale model outperforms the other non-FFT models. The inclusion of the FFT block in the Full model (A) did not affect the F1 score, indicating that the FFT, while increasing the complexity of the model, did not compensate with additional useful information. In the case of the Multiscale + FFT model (E), there {was a marginal decrease in the F1 score, with no statistical significance, over the model (B).} In contrast, the Cropped-16 + FFT model (F) showed significant improvements over Cropped-16 (C) across all metrics, {equal to the best-performing models in terms of F1 score.}

\begin{figure}
\centering
\begin{subfigure}{0.95\columnwidth}
    \centering
    \includegraphics[width=\linewidth]{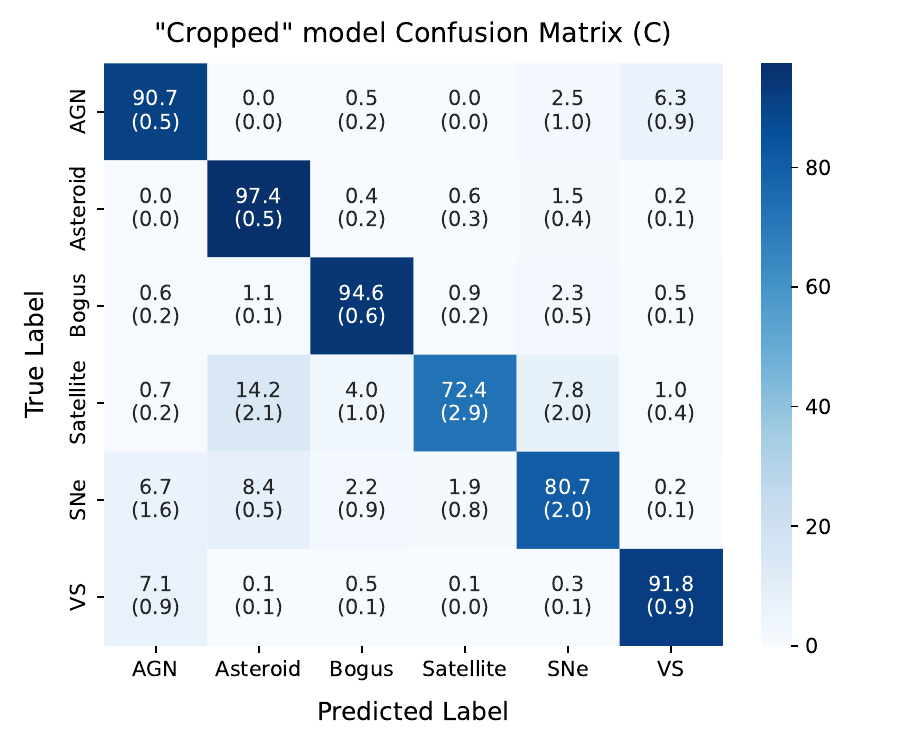}
    \caption{}
    \label{fig:CM_C_cropped}
\end{subfigure}

\vspace{5pt}

\begin{subfigure}{0.95\columnwidth}
    \centering
    \includegraphics[width=\linewidth]{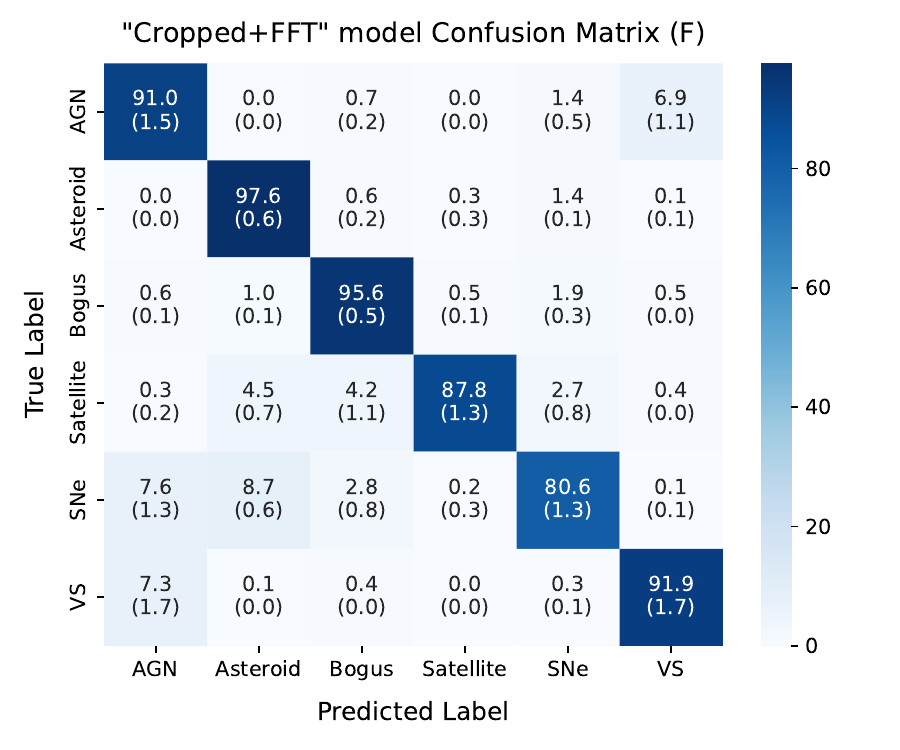}
    \caption{}
    \label{fig:CM_F_cropped_fft}
\end{subfigure}

\caption{Confusion matrices for the Cropped and Cropped + FFT models (C \& F). The values are normalized by the True labels (rows). In parentheses are the standard deviations over 5 runs.}

\label{fig:CMs_cropped}
\end{figure}

\begin{figure}
\centering
\begin{subfigure}{0.95\columnwidth}
    \centering
    \includegraphics[width=\linewidth]{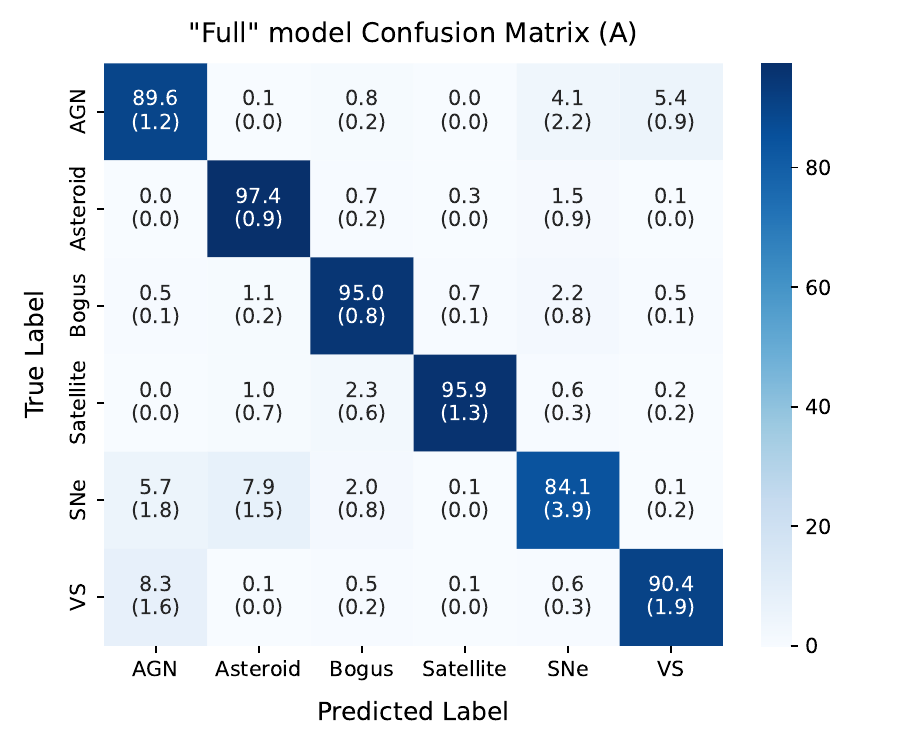}
    \caption{}
    \label{fig:CM_A_full}
\end{subfigure}

\vspace{5pt}

\begin{subfigure}{0.95\columnwidth}
    \centering
    \includegraphics[width=\linewidth]{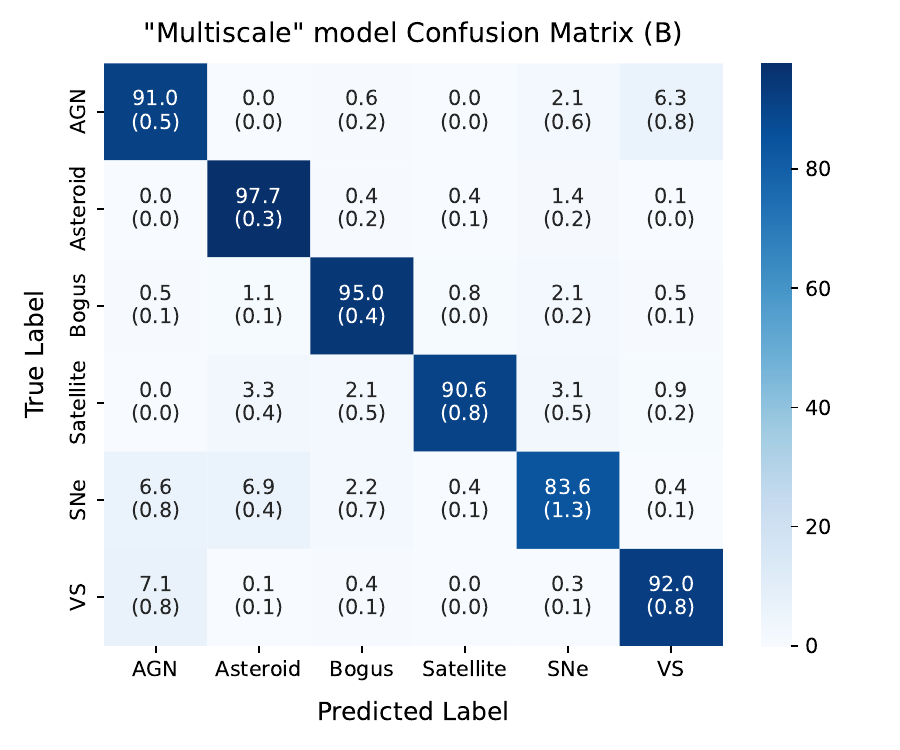}
    \caption{}
    \label{fig:CM_B_multiscale}
\end{subfigure}

\caption{Same as Fig.~\ref{fig:CMs_cropped} for the Full and Multiscale models.}
\label{fig:CMs_base}

\end{figure}

Fig.~\ref{fig:CMs_cropped} shows the confusion matrices for the Cropped-16 (C) and Cropped-16 + FFT (F) models. The addition of the FFT notably reduced misclassifications of satellites, bringing performance {close to} models A and B as noted in Fig.~\ref{fig:CMs_base}. The detailed analysis of these results follows in the next section.

\section{Discussion}
\label{sec:discussion}

We evaluate here the impact of incorporating the FFT in satellite detection and explore potential applications for internal survey pipelines responsible for generating transient alerts. By enhancing the early detection and classification process, FFT-based methods may help streamline the flow of high-quality alerts in large-scale surveys. 
{We first analyze the classification metrics and confusion matrices (Sect.~\ref{sec:discussionFFT}), followed by an assessment of the satellite sample properties, potential biases, and limitations (Sect.~\ref{sec:propSat}), before discussing broader implications for future survey pipelines (Sect.~\ref{sec:discussionBeyond}).}

\subsection{The impact of the spatial frequency space}
\label{sec:discussionFFT}

In Fourier space, the distinct signatures of satellites, characterized by transient and extended features, become clear. Satellite glints, challenging to detect in the spatial domain, are efficiently isolated in the FFT space, enhancing detection accuracy and providing a compact data representation for ML models.

To evaluate the benefits of FFT, we categorize image information into three types (source, context, and satellite), which correspond to different spatial scales and play a critical role in classification tasks. 
Source information {generally captures small-scale, transient, or variable objects such as SNe, AGN, variable stars, and asteroids}, typically spanning only a few arcseconds in size. {Satellites, however, are an important exception, as we discuss separately below.}
Context information provides larger-scale, unchanging environmental details, such as the position relative to nearby galaxies. The scale of this context information can vary; typical galaxy sizes range from a few arcseconds to several tens of arcminutes for extreme cases like the local Pinwheel Galaxy (M101). An important example is SN 2023ixf, discovered by amateur astronomer Koichi Itagaki \citep{Itagaki_2023_sne_amateur_astron}, within M101. Such a large host galaxy can complicate early classification efforts, as methods relying on context information, such as ALeRCE's Stamp Classifier \citep{Carrasco-Davis_2021_ALeRCE}, might miss SNe in such contexts, as happened in this case, where it was mislabeled as bogus. This underscores the importance of a multiscale approach for capturing crucial contextual data. Satellite information, however, is distinct due to its large linear spread and transient nature, making the FFT particularly effective in isolating it.

Our experiment showed that the Cropped-16 model (C) performed worse overall in the classification task compared to the Full (A) or the Multiscale (B) models (see Table~\ref{tab:Metrics}). Since it contains less context information (smaller FoV), this is to be expected. It was particularly bad at identifying satellites, frequently confusing them with asteroids {and SNe} (see {Fig.~\ref{fig:CM_C_cropped} and} below for more details). By including the FFT Block in the Cropped-16 + FFT model (F), these issues were largely resolved, with satellite identification improving {close to the accuracy of the best performing models (see Fig.~\ref{fig:CMs_base})}. The confusion between satellites (true label), and asteroids and SNe was greatly reduced{, despite the reduced context information.}

These results suggest that the FFT of the difference image is highly effective for identifying satellites and capturing the periodic nature of glints, which are challenging to detect in the spatial domain. {However, models incorporating more contextual information from wider FoVs in the science and reference images (such as the Full and Multiscale models) exhibit significantly less confusion in classifying SNe. This evidences that the FFT alone does not fully compensate for the contextual information gap, limiting the performance of the Cropped-16 + FFT model (F).}

It is important to {emphasize} that the dataset is imbalanced, with significantly more bogus objects than satellites. {Despite the efforts to mitigate class imbalance during model training (see Sect.~\ref{sec:dataAndTraining}), some residual effects of this may persist. In this context, comparing models F (Fig.~\ref{fig:CM_F_cropped_fft}) and B (Fig.~\ref{fig:CM_B_multiscale}), the former performs marginally better in Bogus-Satellite confusion (by 0.3\%~$\pm$~0.1\%), while the latter has a marginal edge in Satellite-Bogus confusion (by 2.1\%~$\pm$~1.6\%). Given the substantial class imbalance ---with 12.7 times more bogus alerts than satellites--- these percentages should be interpreted with caution. In both cases, the accuracy for classifying satellites is less than model A (Fig.~\ref{fig:CM_A_full}), which includes more context from the complete stamp data. 
Overall, the 16$\arcsec$ FoV model F, enhanced by the inclusion of FFT, performs marginally below model B in satellite identification. The residual confusion with SNe likely arises from the reduced contextual information in the smaller FoV, which the FFT alone does not fully compensate for. Additionally, confusion with bogus detections can occur in crowded fields, where poor difference-image subtraction of extended features can mimic satellite-like FFT signatures, even when artifacts are not genuinely satellite-related.}

\subsection{Properties of sampled satellites}
\label{sec:propSat}

{To assess potential biases and limitations in our classification methods, we characterize the properties of the detected satellites based on the available alert data, as access to the full images is not feasible in our setup. Specifically, we extract two key parameters: the brightness of the central glint, taken from the \texttt{magpsf} value in the ZTF alert metadata, and an estimated tumbling period derived from the spatial structure of the glints within the alert stamps.}

\begin{figure}
    \centering
    \includegraphics[width=\linewidth]{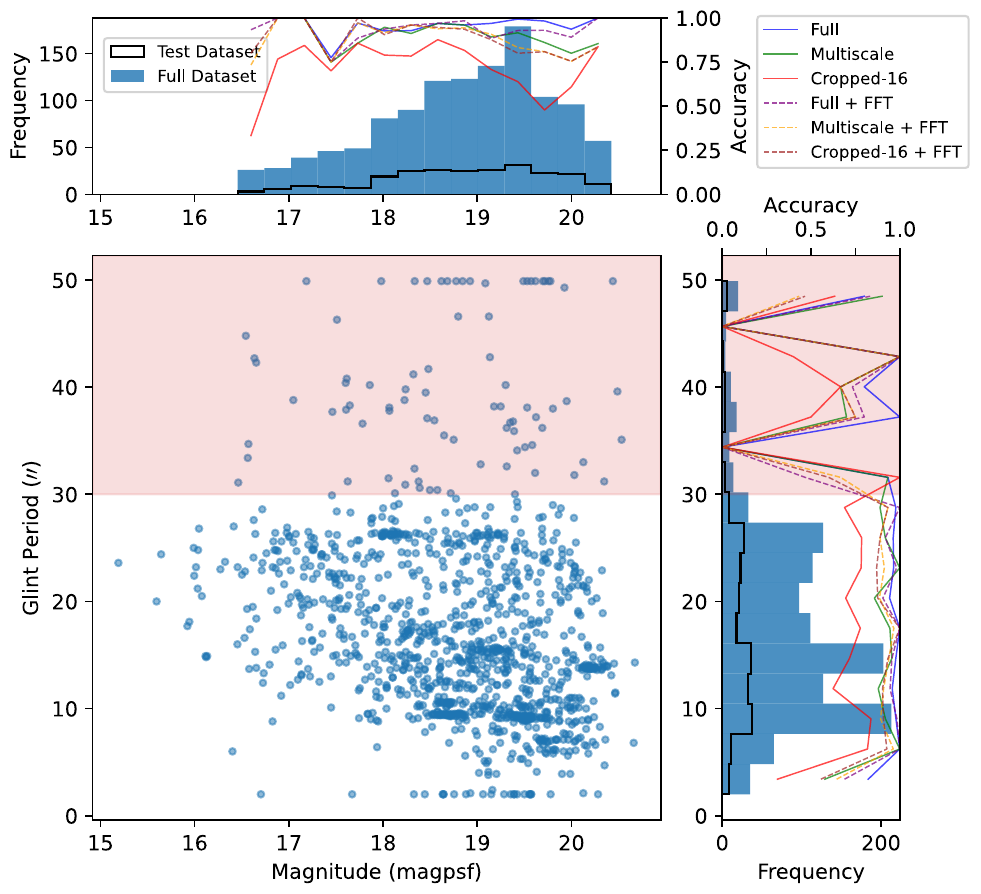}
    \caption{Distribution of magnitudes and estimated glint periods for our satellite dataset. The central scatter plot presents the full sample, with marginal histograms (blue: full sample, black outline: test subset). Colored lines show classification accuracy per model, ranging from 0 (no recovery) to 1 (perfect recovery). Due to the limited 63$\arcsec$ stamp FoV, glint periods $\gtrsim$30$\arcsec$ are incompletely sampled (shaded region, see Sect.~\ref{sec:propSat}). The Full model shows consistently high accuracy, whereas the Cropped-16 model accuracy drops notably at $\sim$19.5 mag. FFT-based models show no glint-period bias, with only minor accuracy reductions at faint magnitudes.}
    \label{fig:propSatPop}
\end{figure}

{The tumbling period is not directly measurable from the alert stamps due to the limited field of view (FoV) of 63$\arcsec$. Given that the ZTF exposure time is $\sim$30 s, a full trail could, in principle, provide a direct measurement of the satellite’s rotation period. However, within an individual alert stamp, we only observe a segment of the satellite’s light trail, imposing a fundamental limitation on how long a periodicity can be reliably measured. As a proxy, we estimate the glint period in arcseconds, i.e., the characteristic spatial frequency of the satellite’s glinting pattern within the stamp. This is obtained by: (i) extracting the light profile along the direction of the trail, (ii) performing a 1D Lomb-Scargle periodogram to estimate the dominant periodicity in arcseconds. Given the stamp FoV, measurable glint periods are significantly limited to $\lesssim$30$\arcsec$ (if aligned with the edges) or $\lesssim$40$\arcsec$ (if diagonal). Longer periodicities arise only in special cases, such as objects displaying secondary glints of different brightness levels (e.g., Fig.~\ref{fig:Sat_signature}, panel e) or irregular, extended trails (e.g., Fig.~\ref{fig:Sat_signature}, panel h). This selection effect is visually represented by the shaded region in Fig.~\ref{fig:propSatPop}.}

{Fig.~\ref{fig:propSatPop} shows the joint distribution of magnitudes and estimated glint periods for our full satellite dataset. The magnitude distribution aligns well with previous studies} \citep[e.g.,][]{Karpov_2022_satellite_glints}{, with a peak around 19.5 mag and a decline at fainter magnitudes. While no prior references exist for the expected glint period distribution, our dataset serves as a useful baseline for evaluating classification biases. The test set (black outline) is representative of the full sample, ensuring that model performance is assessed on a well-sampled population.}

{Overlaid on Fig.~\ref{fig:propSatPop} are the accuracy distributions for each classification model, averaged over the 5 training runs. Model A, with full-resolution and 63$\arcsec$ FoV, shows uniform accuracy close to 1 across all magnitude and glint period bins, confirming that access to the entire stamp preserves satellite properties without bias. The Cropped-16 model (16$\arcsec$ FoV) exhibits significantly lower recovery rates, particularly around 19.5 mag, where a sharp drop is observed. The rest of the models (the Multiscale and the FFT-enhanced) perform similarly across glint periods, showing no systematic bias. However, they exhibit a minor decline in accuracy at fainter magnitudes, decreasing to $\sim$0.8 at 20 mag. This suggests that while the FFT method successfully recovers satellite information, it begins to struggle at lower signal-to-noise ratios when the full context is not available.}

{Importantly, FFT-based approaches show no significant bias in glint-period recovery. This reinforces the suitability of FFT for encoding large-scale features while maintaining sensitivity to periodic satellite glints.}

\subsection{Future Prospects}
\label{sec:discussionBeyond}

\begin{figure*}[!htb]
    \centering
    \includegraphics[width=0.9\textwidth]{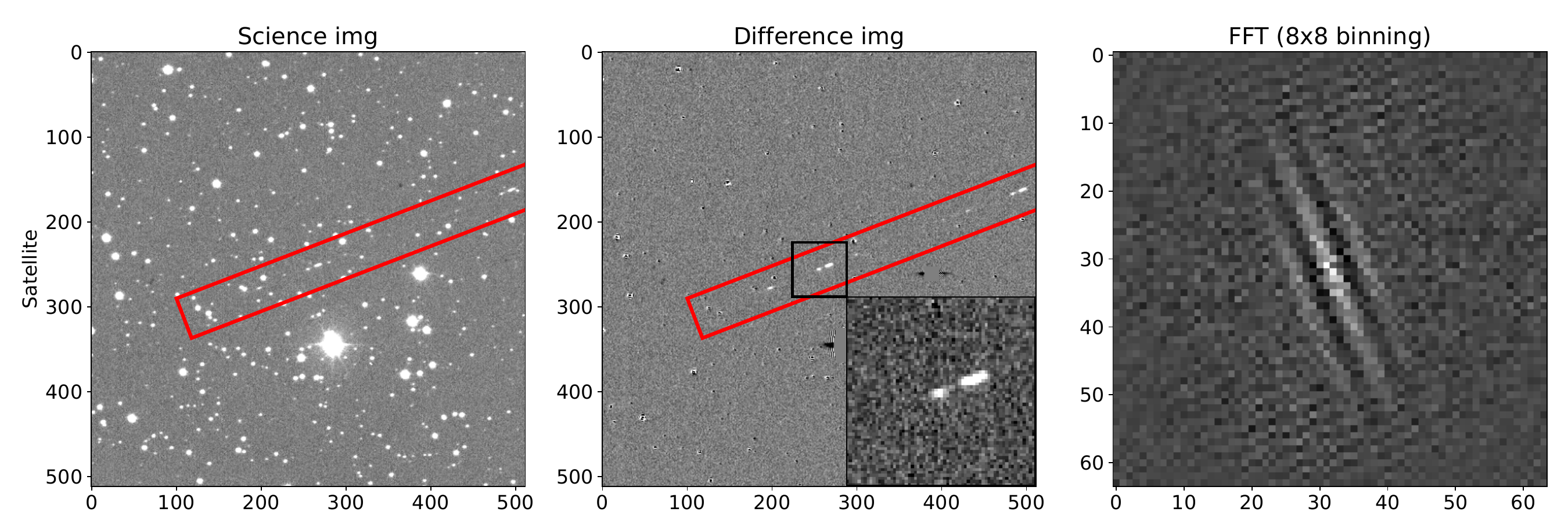}
    \caption{512~$\times$~512 pixel cutouts from a ZTF full-chip image containing the alert-triggering object \texttt{ZTF20aapfawc} at the center. The satellite trail, highlighted with a red box, spans across the cutouts. Additional alerts raised by this satellite streak correspond to objects \texttt{ZTF20aapfavz} and \texttt{ZTF20aapfavx}, located 60$\arcsec$ to the bottom-left and 250$\arcsec$ to the top-right of the center, respectively, along the highlighted trail. The left and center panels display science and difference images, respectively, while the right panel shows the FFT of the difference image with an 8-pixel binning. In spatial frequency space, the satellite exhibits a distinctive fringe-like pattern, similar to the smaller stamps shown in Fig.~\ref{fig:Sat_signature}.}

    \label{fig:FFT_large_cutout}
\end{figure*}

In this study, we focused on the amplitude of the FFT, using it as a single additional channel for data compression. However, since the FFT provides a complex field, the phase information, though not considered here, may also prove valuable in distinguishing certain features. Future work could investigate alternative approaches, such as incorporating 1-channel real-part FFT or 2-channel real- and imaginary-part FFT components, to capture a wider range of spatial frequency features.

Beyond the technical aspects of the FFT, its practical applications in large-scale surveys like LSST are worth exploring. Integrating FFT into internal pipelines could significantly improve the identification of satellite contaminants before the data reaches downstream (alert) processing. 
{However, for the FFT to be most effective, the satellite trail must be well-centered in the image. Our experiments showed that even small offsets (e.g., 5 pixels within the 63$\arcsec\times$63$\arcsec$ stamps) significantly distort the signal, regardless of whether the offset is along the trail direction or cross-trail. Larger offsets quickly cause the signal to disappear entirely. Although, in principle, a shift in the spatial domain corresponds to a phase modulation in Fourier space ($\mathcal{F}[f(x + c)] = F(k) e^{i 2 \pi k c}$), numerical effects in the FFT can degrade the signal. Therefore, we caution against relying on FFT for stamps with substantially offset satellite glints or trails.}

Fig.~\ref{fig:FFT_large_cutout} demonstrates the application of FFT over a larger cutout (512~px), beyond the typical 63~px of ZTF alerts. The central portion of the image focuses on a satellite trail, initially mislabeled as bogus but correctly identified as a satellite by our FFT-enhanced model, later confirmed by human inspection using the ALeRCE Explorer\footnote{\href{https://alerce.online/}{https://alerce.online/}}. Binning the FFT (right panel of Fig.~\ref{fig:FFT_large_cutout}) smooths out high-frequency details, allowing us to focus on large-scale variations that are more characteristic of satellite signatures. This approach demonstrates how FFTs could be implemented in internal pipelines, leveraging the information from very extended trails that produce apparently independent blobs at smaller scales, which may otherwise even fall below the detection threshold. Identifying and intelligently masking out these extended tracks could greatly reduce the number of complex satellite stamps which currently pass into the alert stream.

When comparing our method to other approaches for satellite detection, several distinctions arise. For instance, \citet{Karpov_2022_satellite_glints, Karpov_2023_satellite_glints} used tracklet reconstruction to identify satellites by correlating non-repeating transient sources geometrically. While this is effective, and helped to reveal the impact satellites have on surveys, our method offers the advantage of considering both the bright and low-brightness features of satellite trails, which could become increasingly important as the LSST's deeper exposures reveal fainter satellite glints.
Other approaches for detecting trails, mentioned in Sect.~\ref{sec:intro}, involve analyzing images to search for elongated trail-like shapes \citep[e.g., using the \texttt{CREATETRACKIMAGE} task from][]{Laher_2014_streak_mask} or employing CNNs \citep[][]{Duev_2019_streaks}, similar to this work. 
Unconnected blobs from satellite glints pose a challenge for the former method. Regarding the latter, a promising area for future development might be to incorporate FFTs into it, given the relative ease to include FFT in ALeRCE's Stamp Classifier. 

However, it is important to acknowledge that no single method is likely to completely address the problem of satellite contamination. While we have highlighted the strengths of the FFT approach, reducing the misclassification of satellite events will likely require a combination of techniques, each addressing different aspects of the contamination problem. Combining methods with complementary strengths, such as tracklet reconstruction, trail detection, CNN-based models, and FFT-enhanced classification, should offer more robust mitigation of satellite contamination in large-scale surveys like LSST.

Another promising avenue involves trail-fitting techniques like those used by \citet{Veres_2012_trail_fitting} for asteroids, which could potentially be adapted to include light-curve estimation. This could allow for the subtraction of satellite signatures from images. Such trail-modeling could be crucial for distinguishing between human-made and natural fast-moving objects, as well as broader space pollution characterization.

As of this writing, the public plans for the Rubin Observatory's data management facilities do not explicitly mention the use of graphics processing units (GPUs). However, given their widespread adoption in accelerating ML pipelines, GPUs would likely be a valuable addition to the survey's infrastructure, either from the outset or as a future upgrade. GPU-optimized FFT algorithms, known for their scalability and efficiency \citep[e.g., see][and Appendix~\ref{sec:benchmarking}]{Moreland_2003_fft_gpu_early, nvidia_cufft_2023, amd_rocfft_2023}, offer an effective solution for processing the massive volumes of high-resolution data that LSST will generate. Their capacity for rapid FFT processing is essential to ensure that FFT-based feature extraction keeps up with the data flow in large-scale surveys like LSST {(see Appendix~\ref{sec:benchmarking})}. Incorporating GPUs into LSST's on-site computation facilities would significantly boost processing efficiency and scalability, though adapting existing pipelines to new hardware and software may pose challenges.

\section{Conclusions}

The study of transient astrophysical events in modern surveys such as ZTF and the LSST comes with several challenges, including managing the vast volumes of data and mitigating the presence of contaminants like CCD artifacts, template subtraction artifacts, and satellite trails. 
In the context of the burgeoning space industry, the latter is a rapidly growing problem that has yet to be adequately addressed by time-domain surveys \citep[see, e.g.,][and references therein for a recent review]{Catelan_2023}. The current LSST bandwidth constraints on alert data volume may strongly limit the ability of downstream brokers and users to optimally filter out various contaminants. Thus, investigating efficient methods to improve the identification and filtering of contaminants early in the data stream, or alternatively to compress the data information more effectively into alerts, will hopefully ensure that the most scientifically valuable information is retained and leveraged by end users.

In this study, we explored the utility of the FFT in enhancing the detection of satellite glints within astronomical survey data streams and its potential as an image compression tool.
We specifically utilized ZTF data, while also anticipating the challenges of the upcoming LSST era. Our results demonstrate that the FFT effectively isolates the distinct signatures of satellite trails and glints, concentrating these features in a compact region of Fourier space. This compactness allows for efficient data compression, significantly reducing the amount of non-informative data.

Our experiment compared the performance of ML models using FoVs of different sizes for the three ZTF alert image stamps (science, reference, difference) and assessed the impact of adding a central cutout of the FFT of the difference image as an input. As discussed in Sect.~\ref{sec:discussionFFT}, incorporating the FFT block significantly improved the classification metrics of model C (which adopted a smaller 16$\arcsec$ FoV) to the levels achievable with the full 63$\arcsec$ FoV. In model F, using 16$\arcsec$ FoV stamps that included an FFT image, the overall F1 score increased to the level of the best performing models, with the most notable improvement on the classification of satellites. The confusion between SNe and asteroids, which can be the product of the small FoV, was not alleviated by the inclusion of the FFT.
In this case, we found that the additional contextual information of larger FoVs was important, and the multiscale model introduced in RJ23 (model B here) did a better job while keeping the size of the input low.

Beyond the immediate scope of this work, the FFT's potential applications in internal data processing pipelines are vast. As discussed in Sect.~\ref{sec:discussionBeyond}, integrating FFTs into existing models can enable the analysis of larger image cutouts without a substantial increase in computational cost, facilitating more accurate and scalable contaminant detection{. As shown in Appendix~\ref{sec:benchmarking}, the GPU-based computation of the FFT would allow to extract satellite signatures in 512$\times$512 pixel images, instead of 64$\times$64 pixel, at virtually no runtime cost}. This approach could be particularly impactful for LSST, where efficient data handling and precise contamination filtering will be paramount.

Moving forward, this method opens several avenues for future optimization. For instance, since the signature of satellites in Fourier space is fringe-like, we found success in experiments to reduce the images into 1D arrays (``spectra''). The possibility of deriving features related to this 1D frequency space spectra or even coefficients that could capture the 2D nature may be of relevance as information to include in alert packets in these data streams. Additionally, integrating this approach with real-time data processing pipelines could offer a substantial improvement in the purity and efficiency of alert systems, providing a more robust framework for upcoming large-scale surveys.

In conclusion, the FFT offers a powerful tool for both enhancing contaminant detection and improving data compression in astronomical surveys. By incorporating these techniques into future data processing pipelines, we can better prepare for the challenges of the LSST era and beyond, ultimately enabling more efficient and accurate astronomical research.

\begin{acknowledgements}
We acknowledge support from the National Agency for Research and Development (ANID) grants: Millennium Science Initiative ICN12\_009 (FEB, AMMA, IRJ, MC) and AIM23-0001 (FEB, FF, MC), BASAL Center of Mathematical Modeling Grant FB210005 (FF, AMMA), BASAL projects ACE210002 (AB, MC) and FB210003 (JPC, FEB, AB, MC), FONDECYT Regular 1241005 (FEB), FONDECYT Regular 1231637 (MC), Beca de Doctorado Nacional (JPC). We also acknowledge the use of the Kultrún computing cluster at Universidad de Concepción, funded by Conicyt Quimal \#170001, Anillo ACT172033, Fondecyt regular 1180291, Fondecyt Iniciacion 11170268, Basal AFB-170002, and Núcleo Milenio Titans NCN19-058. {We are grateful to the anonymous referee for their careful review and valuable feedback, which significantly improved the clarity and robustness of this work.}
\end{acknowledgements}

\bibliographystyle{aa}
\bibliography{references}

\begin{appendix}

\section{Computational Benchmarking of FFT Performance}
\label{sec:benchmarking}

\begin{figure}[!htb]
    \centering
    \includegraphics[width=0.9\linewidth]{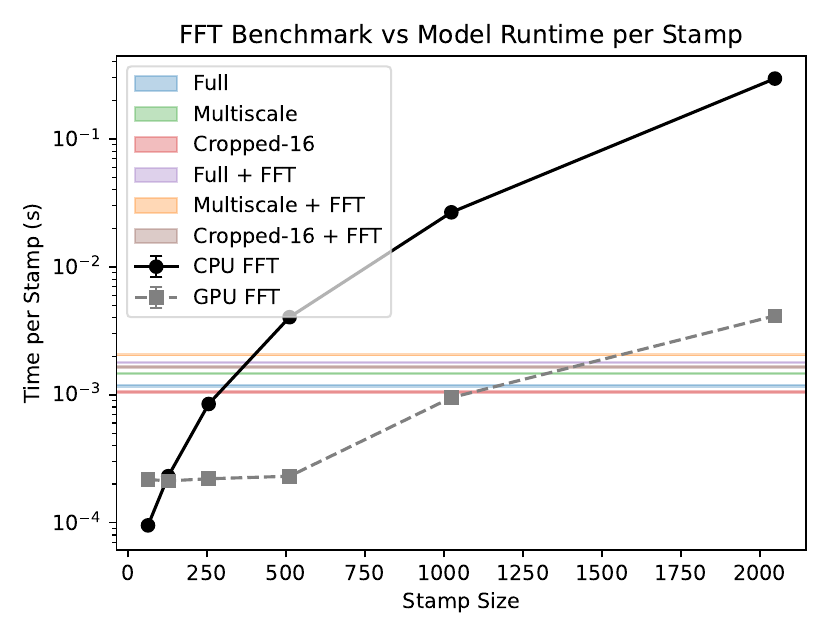}
    \caption{Scaling of 2D-FFT computation times with image size (CPU vs. GPU) and the runtime for each of our models as horizontal spans, with the width being the standard deviation between model runs. We provide the runtime for the models for perspective, but these do not vary stamp size.}
    \label{fig:fft_benchmark}
\end{figure}

To evaluate the feasibility of integrating FFT-based preprocessing into real-time astronomical alert pipelines, we benchmarked the computational cost of FFT computations (both CPU- and GPU-based implementations) and compared it to the runtime of our deep learning models. This analysis provides a direct assessment of whether FFT processing introduces a significant computational overhead and whether it scales efficiently for larger images, such as those expected in upcoming surveys like LSST.

For consistency and reproducibility, both benchmarks were executed on the same computational system: a single NVIDIA RTX A4000 GPU for GPU benchmarks, and a single computational thread of an AMD EPYC 7773X CPU for CPU benchmarks.

\subsection*{FFT Computation Runtimes}

We computed the 2D FFT on square stamps of various sizes ($2^n$, $n\in[6,11]$) using the \texttt{numpy.fft.fft2} function for the CPU-based implementation \citep{Harris2020numpy}, and the \texttt{cupy.fft.fft2} for the GPU-based implementation \citep{cupy_learningsys2017}.

Each FFT computation was repeated 30 times per image size using different random cutouts from a single ZTF difference image to measure execution time and variability. The difference in performance between CPU and GPU implementations is significant: the GPU-based FFT is an order of magnitude faster at 512$\times$512 pixels and two orders of magnitude faster at 2048$\times$2048 pixels compared to CPU execution.

\subsection*{Machine Learning Model Runtimes}

To provide a meaningful baseline for interpreting FFT overhead, we also measured the inference time per alert for each of the six model architectures introduced in Sect.~\ref{sec:models} (with GPU acceleration). For each model, we computed runtimes over five independent training runs, evaluated on the full 9,888-alert test set, and reported the mean and standard deviation. These results are shown as horizontal spans in Fig.~\ref{fig:fft_benchmark}, where the width represents the standard deviation across runs.

FFT computation remains a fraction of the time required for CNN inference. On a GPU, FFT execution is sub-millisecond fast, even for 1024$\times$1024 pixel images.

These results confirm that FFT-based preprocessing is highly efficient and feasible for real-time pipelines of large-scale surveys such as ZTF and LSST.
\end{appendix}
\end{document}